# Measuring Woody: The Size of Debian 3.0[*]

Juan José Amor, Gregorio Robles, and Jesús M. González-Barahona

December 2004

**Abstract**

Debian is possibly the largest free software distribution, with well over 4,500 source packages in the latest stable release (Debian 3.0) and more than 8,000 source packages in the release currently in preparation. However, we wish to know what these numbers mean. In this paper, we use David A. Wheeler's SLOCCount system to determine the number of physical source lines of code (SLOC) of Debian 3.0 (aka woody). We show that Debian 3.0 includes more than 105,000,000 physical SLOC (almost twice than Red Hat 9, released about 8 months later), showing that the Debian development model (based on the work of a large group of voluntary developers spread around the world) is at least as capable as other development methods (like the more centralized one, based on the work of employees, used by Red Hat or Microsoft) to manage distributions of this size.

It is also shown that if Debian had been developed using traditional proprietary methods, the COCOMO model estimates that its cost would be close to $6.1 billion USD to develop Debian 3.0. In addition, we offer both an analysis of the programming languages used in the distribution (C amounts for about 65%, C++ for about 12%, Shell for about 8% and LISP is around 4%, with many others to follow), and the largest packages (The Linux kernel, Mozilla, XFree86, PM3, etc.)

# 1 Introduction

On July 17th of 2002, the Debian Project announced Debian GNU/Linux 3.0 [Debian30Ann] [Debian30Rel]. Code named "woody", it is the latest (to date) release of the Debian GNU/Linux Operating System.

---

[*]This work has been funded in part by the European Commission, under the CALIBRE CA, IST program, contract number 004337.

In this paper, we have analyzed this distribution, showing its size, and comparing it to other contemporary distributions.

Debian was not only possibly the largest GNU/Linux distribution at time of its release, it is also one of the more reliable, and enjoys several awards based on users preferences[1]. Although its user base is difficult to estimate, since the Debian Project does not sell CDs or other media with the software, it is certainly important within the Linux market[2]. It also takes special care to benefit from one of the freedoms that libre software provides to users: the availability of source code. Because of that, source packages are carefully crafted for easy compilation and reconstruction of original (upstream) sources. This makes it also convenient to measure and, in general, to get statistics about it.

We decided to write this paper as an update to *Counting Potatoes* [GBarahona2001], which was written covering the Debian 2.2 release, two years ago. Originally, the idea of these papers came after reading David A. Wheeler's fine paper [Wheeler2001]. We encourage the reader to at least browse it, and compare the data it offers for Red Hat Linux with those found here.

The structure of this paper is as follows. Next section provides some background about the Debian Project and the Debian 3.0 GNU/Linux distribution. After that, we discuss the method we have used for collecting the data shown in this paper. Later on, we offer the results of counting Debian 3.0 (including total counts, counts by language, counts for the largest packages, etc.). The following section offers some comments on the numbers and how they should be understood, and some comparisons with Red Hat Linux and other operating systems, both free and proprietary. To finish, we include some conclusions and references.

## 2 Some background about Debian

The Debian 3.0 GNU/Linux distribution is put together and maintained by the Debian project. In this section, we offer some background data about Debian as a project, and about the Debian 3.0 release.

### 2.1 The Debian project

Debian is a free (libre) operating system, which currently uses the Linux kernel to put together the Debian GNU/Linux software distribution (although other distributions based in other kernels, like the Hurd and FreeBSD, are now in pre-testing phase and expected in the near future). This distribution is available for several architectures, including Intel x86, ARM, Motorola 680x0, PowerPC, Alpha, and SPARC.

The core of the Debian distribution (called section "main", which amounts for the vast majority of the packages) is composed only of free software. It is available on the Net for download, and many redistributors sell it in CDs or other media. The Debian distribution is put together by the Debian project, a group of over 900 volunteer developers spread around the world, collaborating via the Internet. The work done by those developers includes adapting and packaging all the software included in the distribution, maintaining several Internet-based services (web site, on-line archive, bug tracking system, support and development mail lists, etc.), several translation and internationalization efforts, development of tools specific to Debian, and in a wide sense, of all the infrastructure that makes the Debian distribution possible.

---

[1] A non exhaustive list of those awards can be found at <http://www.debian.org/misc/awards>

[2] Just as an example, the Netcraft study about the web servers in the whole Internet found on January 2004 that Debian was the fastest growing Linux distribution, more than 24% during that year, and the third in absolute figures, see <http://news.netcraft.com/archives/2004/01/28/debian_fastest_growing_linux_distribution.html>.

Debian developers package software which they obtain from the original (upstream) authors, ensuring that it works smoothly with the rest of the programs in the Debian system. For this matter, there is a set of rules that the package should comply with, the so-called Debian Policy Manual [DebianPol]. Most of the work for packaging a given program is usually to make it compliant with those rules. Developers also take bug reports, try to fix them (reporting fixes and problems upstream), follow new upstream developments, and build all the software glue needed for making the Debian system work. Bugs and security holes are discussed openly, and updates fixing important problems are made available for stable releases on a daily basis so that users can maintain their systems secure and as bug-free as possible.

Debian is unique for many reasons. Its dedication to free software, its non-profit nature, and its open development model (where most of the discussions are addressed openly in public lists) are remarkable. The project is committed to free software, as is reflected in the Debian Social Contract (see [DebianSocialContract]). The definition of what Debian understand as free software can be found in the Debian Free Software Guidelines ([DFSG]), and in essence is the same software which can be considered "open source" software (which is not strange, since the Open Source Definition was actually derived from the DFSG).

## 2.2  Debian woody

Debian 3.0 (woody) is the latest official release, the one currently considered as "stable". It was released on July of 2002, and includes all the major libre software packages available at that time. Only in its main distribution, composed only of free software (according to the Debian Free Software Guidelines), there are more than 4,500 source packages. The whole release includes almost 10,000 binary packages, which the user can install easily from various media or from Internet servers.

In addition to woody (the stable release), the Debian archives include two releases in preparation: "testing" and "unstable". Currently, testing is codenamed "sarge", and is in freezing process. In the next months it will be released as Debian 3.1. The unstable release (currently "sid") is used by developers to upload the packages, and test them. If during a certain period there are no release critical bugs[3] filed against them nor against the packages on which they depend, they are passed to testing, and are considered for the following stable release.

Debian 3.0 is split in two archives: the "regular" one, and the "non-US" archive. In non-US are archived those packages which have some legal impediment to be exported from the United States of America (usually the the US legislation on strong cryptography). Non-US is currently being deprecated, as software related to cryptography is entering the regular archive. It is expected that Debian 3.1 will not include non-US.

Each archive is composed of several so called "distributions": **main**, **contrib** and **non-free**.

For the work referenced in this paper we have considered only the **main** distribution of the "regular" archive. It is the largest (by far) fraction of the archive, is only composed of free software, and has no export restrictions. In many respects, it is one of the largest coordinated collection of free

---

[3]Each bug in Debian is assigned a severity (see the list of severity types in <http://www.debian.org/Bugs/Developer>. Some of the types are considered "release critical", meaning that the affected package will not enter the new stable release until it is fixed.

software available in the Internet.

# 3 Collecting the data

The approach used for collecting the data presented in this paper is, in summary, as follows:

1. Which source code makes a Debian release?

   Fortunately enough, source code for current and past Debian releases is archived, and available for everyone in the Internet. The only problem is to determine the list of source packages for any given release, and where to access them.

2. Downloading and collecting data

   Once we know what files to download, we have to download all of them before being able of gathering data. Since the size of the unpackaged sources for a Debian release is very large, we chose to work on a per-package basis, gathering all the relevant data from it (mainly, the number of lines of code) before deleting it and following on with the download of the next one.

3. Final analysis

   Analyze the collected data and get some statistics regarding the total number of physical SLOC of the release, the SLOC for each package, the SLOC for each of several programming languages considered, etc.

In the following sections these three steps are described in more detail.

## 3.1 Which source code makes a Debian release?

The Debian packaging system considers two kind of packages: source and binary. One or more binary packages can be built automatically from each source package. For instance, from a certain source package maybe three different binaries packages are built: one with a library, another with and executable program and a third one with documentation. For this paper, only source packages are relevant, and therefore we will no longer refer to binary packages.

When building a source package, a Debian developer starts with the "original" source directory for the piece of software. In Debian parlance, that source is called "upstream". The Debian developer patches upstream sources if needed, and creates a directory **debian** with all the Debian configuration files (including data needed to build the binary package). Then, the source package is built, usually (but not always) consisting of three files: the upstream sources (a **tar.gz** file), the patches to get the Debian source directory (a **diff.gz** file, including both patches to upstream sources and the **debian** directory), and a description file (with extension **dsc**). Patches files are not present for "native" source packages (those developed for Debian, with no upstream sources).

Source packages of current Debian releases are part of the Debian archive. For every release, they reside in the **source** directory. There are sites in the Internet including the source packages for every official Debian release to date (usually, mirrors of archive.debian.org <`ftp://archive.debian.org`>). Since Debian 2.0, for every release a **Sources.gz** file is present in the **source** directory, with information about the source packages for the release, including the files that compose each package. This is the information we use to determine which source packages, and which files, have to be considered for Debian 3.0.

However, not all packages in **Sources.gz** should be analyzed when counting lines of code. For instance, in some cases, there are several versions of the same piece of software. As an example, in Debian 3.0 we can find source packages **gcc272**, **gcc2.95**, **gcc2.96** and **gcc3.0**. Counting all of these packages will imply counting GNU C Compiler four times, which is not the intended procedure. Therefore, a manual inspection of the list of packages is needed for every release, detecting those which are essentially versions of the same software, and choosing one "representative" for each family of versions.

These cases may cause an underestimation of the number of lines of the release, since different versions of the same package may share a lot of code, but not all (consider for instance PHP4 and PHP3, with the former being an almost complete rewrite of the latter). However, we think this effect is negligible, and compensated with some overestimations (see below).

In other cases, we have decided to analyze packages which may have significant quantities of code in common. This is the case, for instance, of **emacs** and **xemacs**. Being the latter a code fork of the former, both share a good quantity of lines which, even when not being exactly equal, are evolutions of the same "ancestors". Other similar case is **gcc** and **gnat**. The latter, an Ada compiler, is built upon the former (a C compiler), adding many patches and lots of new code. In those cases, we have considered that the code is different enough to consider them as separate packages. This probably leads to some overestimation of the number of lines of code of the release.

The final result of this step is the list of packages (and the files composing them) that we consider for analyzing the size of a Debian release. This list is done by hand (with the help of some really simple scripts) for each release.

## 3.2 Downloading and collecting data

Once the packages and files composing Debian 3.0 are determined, they are downloaded from some server of the net of Debian mirrors. Some simple Perl scripts where used to automate this process, which (for each package) consists of the following phases:

- Downloading of the files composing the package

- Extraction of the source directory corresponding to the upstream package (by untaring the **tar.gz** file. After extraction, data about this upstream source is gathered.

- Patching of the upstream directory with the **diff.gz** file, to get the Debian source directory. After extraction, data about it is gathered.

- Deletion of the **debian** directory, to avoid counting maintainer scripts (stored in this directory), and gathering data about this sans-debian Debian source package.

Not all packages have upstream version. Therefore, during this process, some care has to be taken to differentiate this situations.

The fetching of data is done using **SLOCCount** scripts (see [SLOCCount]), three times for each package (one in each phase, see above), which stores the count of lines of code for each package in a separate directory, ready for later inspection and reporting.

The reason for fetching data three times for every package is to analyze the impact of the Debian developer on the source package. This impact can be in the form of patches to the source (usually to make it more stable and secure, to conform to Debian installation policy, or to add some functionality to it) or in installation scripts (which can be singled out when counting sans-debian source packages).

The final result of this step is the collection of all the data fetched from the downloaded packages, organized by package, and ready to be analyzed. These data consist mainly of lists of files and line counts for them, split by language.

## 3.3 Final analysis

The last step is the generation of reports, using **SLOCCount** and some scripts, to study the gathered data. Since in this step all the fetched data is available locally, and in a simple to parse form, the analysis can be done pretty quickly, and can be repeated easily, looking for different kinds of information.

The final result of this step is a set or reports and statistical analysis, using the data fetched in the previous step, and considering them from different points of view. These results are presented in the following section.

# 4 Results of counting Debian

The main results of our current analysis of the Debian 3.0 GNU/Linux release can be organized in the following categories:

- Size of Debian woody.

- Importance of the most used programming languages.

- Analysis of the evolution in the size of the most relevant packages.

- Effort estimations.

## 4.1 Size of Debian woody

We have counted the number of source lines of code of Debian GNU/Linux 2.2 in three different ways, with the following results (all numbers are approximate, see [DebianCounting] for details):

- Count of upstream packages "as such": *98,400,000* SLOC

- Count of Debian source packages: *105,500,000* SLOC

- Count of Debian source packages without **debian** directory: *105,000,000* SLOC

For details on the meaning of each category, the reader may revisit the subsection "Downloading and collecting data". In short, the count of upstream packages could be considered as the size of the original software used in Debian. The count of Debian source packages represents the amount of code actually present in the Debian 3.0 release, including both the work of the original authors and the work of Debian developers. This latter work includes Debian-related scripts and patches. Patches can be the work of Debian developers (for instance to adapt a package to the Debian policy) or be the downloaded from elsewhere. The count of Debian packages without the **debian** directory excludes Debian-related scripts, and therefore is a good measure of the size of the packages as they are found in Debian, excluding the specific Debian-related scripts.

It is also important to notice that packages developed specifically for Debian have usually no upstream source package. This is, for instance the case of **apt**, which is present only as a Debian source package.

## 4.2 Programming languages

The number of physical SLOC, classified by programming language, are (roughly rounded) as follows (numbers for Debian source packages):

- C: *66,550,000* SLOC (63%)

- C++: *13,000,000* SLOC (12.4%)

- Shell: *8,630,000* SLOC (8.1%)

- LISP: *4,080,000* SLOC (3.9%)

- Perl: *3,200,000* SLOC (3%)

- FORTRAN: *1,900,000* SLOC (1.8%)

- Python: *1,450,000* SLOC (1.4%)

- Assembler: *1,300,000* SLOC (1.3%)

- Tcl: *1,080,000* SLOC (1%)

- PHP: *648,000* SLOC (0.62%)

- Ada: *576,000* SLOC (0.55%)

- Modula3: *571,000* SLOC (0.55%)

Below 0.5% we find some other languages: Java (0.5%), Objective C (0.4%), Yacc (0.37%), ML (0.27%), Lex (0.13%), and others below 0.1%.

In Figure 1 we can view the importance of main languages versus the rest. It is consistent with the fact that most packages are written in C. C++ is another language present in most packages, and main language in some of them (as Mozilla does). The same fact occurs with Lisp, which is the main language in several packages (such as Emacs), but also is used in most of rest packages. Next, shell scripts are used to support configuration and other auxiliary task in most packages.

When we count the lines in the Debian source packages without the **debian** directory (which contains package configuration files and maintainer scripts), the numbers are similar. This means that the maintainer scripts are not a significant part of the distribution. The main difference is in Shell and Perl lines, which uncovers the preferred languages for those scripts.

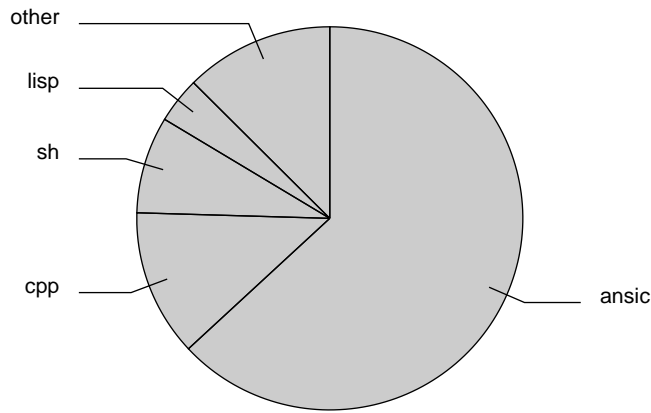

Figure 1: Pie with the distribution of source lines of code for the majoritary languages in Debian 3.0

However, when we count original (upstream) source packages there are some remarkable differences: about 2,000,000 lines of C code, 300,000 lines of LISP, 200,000 lines of FORTRAN, and minor variations in other languages. This differences can usually be amounted to patches to upstream packages made by the Debian developer. Therefore, looking at this numbers, we can know in which languages are written the most patched packages.

## 4.3 The largest packages

The largest packages in the Debian woody distribution are:

- Linux kernel (2.4.18): 2,574,000 SLOC. C amounts for 2,558,000 SLOC, Makefiles, assembler and scripts in several languages amounts for the rest. This is the latest kernel included in Debian 3.0 release.

- Mozilla (1.0): 2,362,000 SLOC. C++ amounts for 1,464,000 SLOC, C for 809,000. Mozilla is the well known open source WWW browser.

- XFree86 (4.1.0): 1,928,000 SLOC. Mainly 1,833,000 SLOC of C. This is an X Window implementation, including graphics server and basic programs.

- PM3 (1.1.15): 1,501,000 SLOC. 891,000 SLOC of C, 545,000 of Modula3, the rest are scripts and some other, mainly C++ code. PM3 is the Modula-3 distribution of the Ecole Polytechnique de Montreal, including a compiler and libraries.

- MingW32 (2.95): 1,500,000 SLOC. Mainly, C code, with 1,137,000 SLOC. MingW32 is the "Minimalist" GNU Win32 cross compiler.

- BigLoo (2.4): 1,065,000 SLOC. Mainly, C code, with 948,000 SLOC and LISP (96,000 SLOC). BigLoo is a Scheme compiler.

- GDB (5.2): 986,000 SLOC. Mainly of C, 904,000 SLOC. This is the well known GNU debugger.

- crash (3.3): 968,000 SLOC. Mainly of C, 890,000 SLOC. It is a kernel debugging utility.

- OSKit (0.97): 921,000 SLOC. Mainly of C, 912,000 SLOC. It is a set of utilities for ease kernel development.

- NCBI libraries (6.1): 831,000 SLOC. Mainly of C, 828,000 SLOC. It is a set of libraries related with biology application development.

Numbers are approximate number of SLOC of the Debian source packages. Only data for the more relevant languages found in each package are reported.

The release numbers of the packages are obviously not current, but those were the ones available at the time of the freeze for Debian 3.0 (Spring 2002). The classification could be different had Debian developers packaged things in other ways. For instance, if all Emacs extensions were in the Emacs package, it would have been much larger (and it have been in this "top ten" list). However, a Debian source package usually matches well with what upstream authors consider as a package, and with the general idea about what is a package.

The next packages by SLOC size (between 807,000 and 600,000 SLOC) are Insight (a graphical debugger), Emacs21 (the well known text editor), Gnat (the GNU Ada95 compiler) and QTEmbeddedFree (the QT Embedded GUI library). The main language for them is C, however, in Emacs we'll find a lot of LISP code, and a lot of Ada in Gnat compiler as expected.

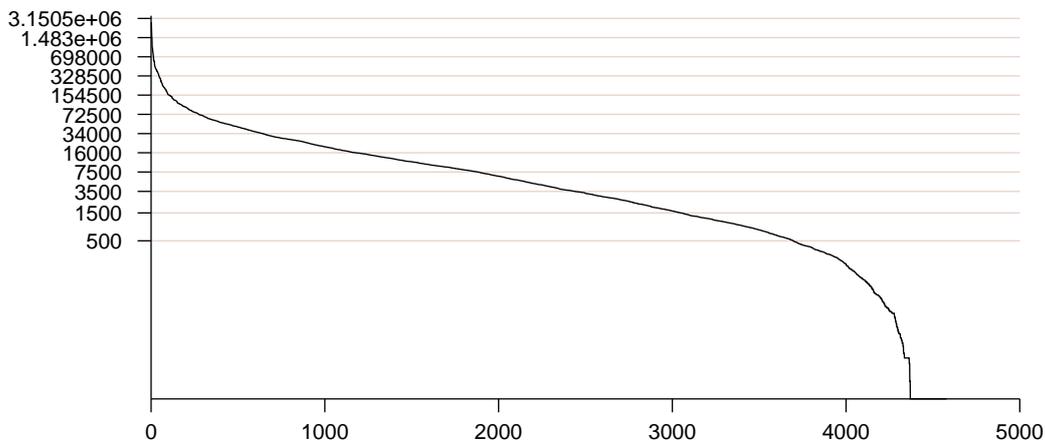

Figure 2: Package sizes for Debian distributions. Packages are ordered by their size along the X axis, while the counts in SLOCs are represented along the Y axis (in logarithmic scale).. Debian 3.0

Figure 2 and Figure 3 show the distribution of size in all packages for Debian 3.0. It can be graphically observed a mean size, that we can measure as 22,860 bytes in Debian 3.0. It is interesting to note that this size, around 23,000, repeats in all Debian releases, from 2.0 to 3.1 (see [Robles2004]).

## 4.4  Effort and cost estimations

Using the basic COCOMO model [Boehm1981], the effort to build a system with the same size as Debian 3.0 can be estimated. This estimation assumes a "classical", proprietary development model, so it is not known if this validly estimates the amount of effort which has actually been applied to build this software. But it can give us at least an order of magnitude of the effort which would be needed in case a proprietary development model had been used.

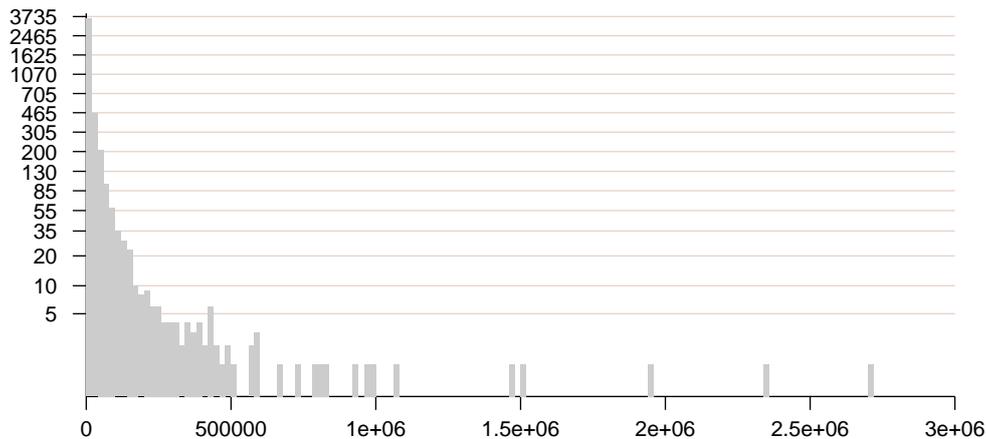

Figure 3: Histogram with the SLOC distribution for Debian packages. Debian 3.0

Using the SLOC count for the Debian source packages, the data provided by the basic COCOMO model are as follows:

- Total physical SLOC count: 104,679,026

- Estimated effort: 322,031.16 person-months (26,835.93 person-years)

  Formula: *2.4 * (KSLOC**1.05)*

- Estimated schedule: 81.72 months (6.81 years)

  Formula: 2.5 * (Effort**0.38)

- Estimated cost to develop: 3,625,000,000 USD

To get these figures, each project was estimated as though it was developed independently from the others, which in nearly all cases is true. For calculating the cost estimation, we have used the mean salary for a full-time systems programmer during 2000, according to Computer World [ComWorld2000], which is of 56,286 USD per year, and an overhead factor of 2.4 (for an explanation on why this factor, and other details of the estimation model, see [Wheeler2001]).

## 5  Some comments and comparisons

The numbers offered in the previous section are no more than estimations. They can give us at least orders of magnitude, and allow for comparisons. But they should not be taken as exact data, there are too much sources of error and field for interpretation. In this section, we will discuss some of the more important assumptions made, and the possible sources of error. We will also compare the SLOC counts with the SLOC counts for other system, with the aim of giving the reader some context to interpret the numbers.

## 5.1 What is a source line of code

Since we rely on David A. Wheeler's **SLOCCount** tool for counting physical SLOC (see [SLOCCount]), we also rely on his definition for "physical source lines of code". Therefore, we could say that we identify a SLOC when **SLOCCount** identifies a SLOC. However, **SLOCCount** has been carefully programmed to honor the usual definition for physical SLOC: "*A physical source line of code is a line ending in a newline or end-of-file marker, and which contains at least one non-whitespace non-comment character.*"

There is other similar measure, the "logical" SLOC, which sometimes is preferred. For instance, a line written in C with two semicolons would be counted as two logical SLOC, while it would be counted as one physical SLOC. However, for the purposes of this paper, the differences between both definitions of SLOC are not that important, specially when compared to other sources of error and interpretation.

## 5.2 Sources of inaccuracy in the SLOC counts

The counts of lines of code presented in this paper are no more than estimations. By no means do we imply that they are exact, specially when they refer to aggregates of packages. There are several factors which cause this inaccuracy of the numbers, some due to the tools used to count, some others due to the selection of packages:

- Some files may have not being counted accurately.

  Although **SLOCCount** includes carefully designed heuristics to detect source files, and to distinguish source lines from comments, those heuristics do not always work as expected. In addition, in many cases it is difficult to distinguish automatically generated files (which should not be counted), although **SLOCCount** makes also a good effort to recognize them.

- Not all programming languages are recognized.

  To fetch the data we used release 2.08 of **SLOCCount**, which recognizes about 20 different languages. However, some languages present in Debian (as is the case of Erlang) are not currently supported. This obviously leads to some underestimation in the packages with files written in those languages.

- Different perceptions in the aggregation of package families and the selection of a representative.

  As we comment in the subsection where we discuss the selection of the list of packages to count, the reasons to take a given package in or out of the list are not unquestionable. Should we count different releases of the same package? Should we count only once code present in several packages, or not? The usual criteria for measuring SLOC is "delivered source lines of code". From this point of view, all packages should be considered as they appear in the Debian release. However, this is difficult to apply when some packages are clearly evolutions of other packages. Instead of considering all of them as "delivered", it seems more productive to consider the older ones as "beta releases". However, in the libre software world it is rather common to deliver stable releases every 6 or 12 months. Those stable releases have a lot of work behind them, only to ensure stability, even if they are also the foundation for later releases.

In most cases, we have adopted an intermediate decision: to count only once families of packages which are a line of evolution (as is the case of **emacs19** and **emacs20**, but to count separately families of packages which happen to share some code but are in themselves different developments (as is the case of **gcc** and **gnat**).

## 5.3 Estimation of effort and cost

Current estimation models, and specifically COCOMO, only consider classical, proprietary development models. But libre software development models are rather different, and therefore those models may not be directly applicable. That way, we can only estimate the cost of the system, had it been developed using classical development models, but not the actual cost (in effort or in money) of the development of the software included in Debian 3.0.

Some of the differences that make it impossible to use those estimation models are:

- Continuous release process (frequent releases). The COCOMO model is based around the concept of "delivered SLOC", which implies one point in the history of the project where the product is released. From there on, the main development effort is devoted to maintenance. On the contrary, most libre software projects deliver releases so frequently that it could be considered as a continuous release process. This process implies the almost continuous stabilization of the code, at the same time that it evolves. Free software projects are used to improve and modify their software at the same time that they prepare it for end users.

- Bug reports and fixes. While every proprietary software system needs expensive debugging cycles, libre software can count on the help of people external to the project, in the form of valuable bug reports, and even fixes for them.

- Reuse, evolution, and inter-fertilization of code. It is common in libre software projects the reuse of code of other libre software projects as an integral part of the system being developed. It is also common that several projects develop evolutions of the same base system, in many cases with all of them using code of the others all the time. Some of this cases can also happen in proprietary developments, but even in large companies, with many open projects, they are not common, while they are the norm in libre software projects.

- Distributed development model. Although some proprietary systems are developed by geographically distributed teams, the degree of distributed development found in libre software projects is several orders of magnitude greater. There are exceptions, but usually libre software projects are carried out by people from different countries, not working for the same company, devoting different amount of effort to the project, interacting mainly through the Internet, and in most cases (specially in large projects), the developer team have never been physically together.

Some of these factors increase the effort needed to build the software, while some others decrease it. Without analyzing in detail the impact of these (and other) factors, the estimation models in general, and COCOMO in particular, may not be directly applicable to libre software development.

## 5.4 Comparison with size estimations for other systems

To put the numbers shown above into context, here we offer estimations for the size of some operating systems. Note that these comparisons are always difficult, and specially when we compare with proprietary systems.

As reported in [Lucovsky2000] (for Windows 2000), [Wheeler2001] and [GBarahona2004] (for Red Hat Linux), and [Schneier2000] and [McGraw2003] (for the rest of the systems), this is the estimated size for several operating systems, in lines of code (all numbers are just approximations):

- Microsoft Windows 3.1: 3,000,000
- Sun Solaris: 7,500,000
- Microsoft Windows 95: 15,000,000
- Red Hat Linux 6.2: 17,000,000
- Microsoft Windows 2000: 29,000,000
- Red Hat Linux 7.1: 30,000,000
- Microsoft XP: 40,000,000
- Red Hat Linux 8.0: 50,000,000
- Debian 2.2: 55,000,000
- Debian 3.0: 105,000,000

Most of these estimations (in fact, all of them, except for Red Hat Linux) are not detailed, and is difficult to know what they consider as a line of code. However, the estimations should be similar enough to SLOC counting methods to be suitable for comparison.

Note also that, while both Red Hat and Debian include many applications, in a lot of cases even several applications in the same category, both Microsoft and Sun operating systems are much more limited in this way. If the more usual applications used in those environments were counted together, their size would be much larger. However, it is also true that all those applications are not developed neither put together by the same team of developers, as is the case in Linux-based distributions.

From these numbers, it can be seen that Linux-based distributions in general, and Debian 3.0 in particular, are some of the largest pieces of software ever put together by a group of developers.

# 6 Conclusions and related work

In this paper, we have presented some results of our work on counting the number of SLOC of Debian 3.0. They represent the state of the Debian GNU/Linux distribution around the summer of 2002. Our estimation is that it amounted for more than 105,000,000 SLOC around that time. Using the COCOMO model, this implies a cost (using traditional, proprietary software development models) close to 3,625 million USD and effort of more than 26,800 person-years. The list of the largest packages, and an analysis by programming language used are also offered. These figures do not even include components such as OpenOffice.org, which were not in Debian 3.0 but are included in the upcoming Debian 3.1.

We can also compare this count to that of other Linux-based distributions, notably Red Hat. Roughly speaking, Debian 3.0 is about twice the size of Red Hat 8.0, which was released about two months later. It is also larger than the latest Microsoft operating systems (although, as is discussed in the corresponding section, this comparison could be misleading).

When coming to the details, some interesting data can be shown. For instance, the most popular language in the distribution is C (more than 60%), followed by C++ (close to 12%), Shell (about 8%), LISP (about 4%), Perl (around 3%) and FORTRAN and Python (about 2%). The largest packages in Debian 3.0 are Linux Kernel (about 2,574,000 SLOC), Mozilla (about 2,362,000 SLOC), XFree86 (more than 1,900,000), and PM3 (1,501,000).

There are not many detailed studies of the size of modern, complete operating systems. Of them, the work by David A. Wheeler, counting the size of Red Hat 6.2 and Red Hat 7.1 is the most comparable. Another interesting paper, with some intersection with this paper is [GodfreyTu2000], a study on the evolution over time of the Linux kernel. Some other papers, already referenced, provide total counts of some Sun and Microsoft operating systems, but they are not detailed enough, except for providing estimations for the whole of the system.

Finally, we find it important to repeat once more that we are offering only estimations. However, we believe they are accurate enough to draw some conclusions and to compare with other systems.

# 7 Acknowledgements

The authors wish to thank Martin Michlmayr, currently the leader of the Debian project, for his comments and suggestions.